\documentclass[preprint,number]{elsarticle}
\usepackage{graphicx}%
\usepackage{dcolumn}%
\usepackage{bm}%

\begin{document}
\title{One-dimensional Modelling of Electrostatic Generation and Detection of Bulk Acoustic Waves 
in layered structures}
\author{Lilia Arapan}
\author{Mihaela Ivan} 
\author{Bernard Dulmet\corref{cor1}}
\ead{bernard.dulmet@femto-st.fr}
\cortext[cor1]{Corresponding author}
\address{FEMTO-ST Institute,
Universit\'e Bourgogne Franche-Comt\'e,\\
ENSMM, 26 rue de l'\'Epitaphe,  25000  Besan\c con, France.}

\begin{abstract}
This paper is devoted to the unidimensional analysis of a 2-ports
silicon resonator vibrating in
thickness--extensional  mode. Both excitation and detection ports are
capacitive transducers used to control the system of longitudinal
elastic  waves established along the thickness of a layered structure.
The analysis consists of integrating the capacitive transduction of
longitudinal elastic waves
 within a specific implementation of the general scheme of impedance
 methods largely used as standard tools for the modelling of  RF-MEMS.
\end{abstract}

\maketitle

\section{Position of problem}
\begin{figure}
\includegraphics[width=0.7\linewidth]{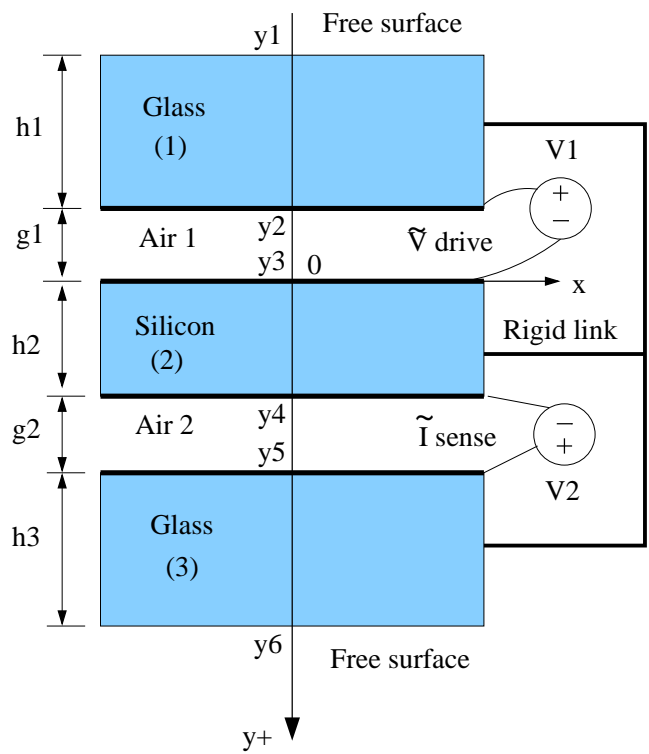}
\caption{Schematic view of the structure under study.}
\label{fig:device}
\end{figure}

This short paper is devoted to the one--dimensional analysis of a silicon
resonator vibrating in
thickness--extensional  mode with a  2--ports capacitive system for  excitation
and detection of the elastic waves. The device under study is schematically represented on
Fig.~\ref{fig:device}. 
The three layers of solid materials form a rigid assembly, leaving 
thin gaps  $g_1$ and $g_2$ between the pairs of inner interfaces. 
Throughout the paper, we use bars and tilde to denote static and dynamic
quantities, respectively. With these conventions,
a driving voltage 
\begin{equation} V_1=\overline V_1+\widetilde V_1 \cos \omega t
\end{equation}
is applied across the first air gap
(input port 1), whereas a purely static voltage
$V_2  \equiv \overline V_2$ is applied across the second air gap (output port 2).
The thickness of the layered structure is oriented along the second axis $\vec
y$ axis of the frame. The extension of the stacked structure along $\vec x$
and $\vec z$ is assumed large enough to justify the one-dimensionnal
assumption, \textit{i.e.} all quantities of interest only depend on the $y$
coordinate. Each air-solid inner interface is coated with an infinitely thin
conductive electrode, and the thickness $h_2$ of the silicon layer is considered
sufficient to neglect electrostatic influence between the electrodes located
at $y_3$ and $y_4$. 

Here-proposed analysis essentially follows a specific implementation of  the general scheme of the
impedance matrices approach \cite{Mason58}, widely used to describe linear problems in the domains of
ultrasonics and piezoelectricity. Since electrostatic excitation of MEMS
devices is intrinsically non--linear, the very first step must consist in linearizing the excitation
problem around the static bias point imposed by $\overline V_1$ and $\overline
V_2$ prior integrating the whole problem within
the formalism of laminar plate Green's functions closely related to acoustic
admittance matrices. The instantaneous
value of the net electrostatic
surface force  is given by:
\begin{equation}
\vec f_e(t)=\frac{\varepsilon_0}{2}\,\frac{ \left[{V_1(t)}\right]^2 }{\left[g_1-u_2(y_2)+u_2(y_3)\right]^2}\,\vec n,
\end{equation}
where $u_2$ denotes the total (static + dynamic) out-of-plane component of the
mechanical displacement along the interface, and $\vec n$ is the  unit
normal to the considered interface, outwardly oriented with respect to the
solid. Since no fringing effect can be accounted
for in the framework of the purely unidimensionnal analysis, the thickness
extensional stress is identical on both sides of the thin electrostatic
gap. With present notations, one obtains:
\begin{equation}\label{eq:total-drive} \displaystyle
T_2(y_2)=T_2(y_3)=\frac{\varepsilon_0}{2} \frac{{
    V_0}^2 + 2 \overline V_1 \widetilde  V_1 \cos \omega t  
           + \frac{{\widetilde
    V_1}^2}{2}\cos 2\omega t   }
{\left(g_1+u_{32}\right)^2},
\end{equation}
where we introduced the notations:
\begin{equation}\begin{array}{ccc} { V_0}=\sqrt{{\overline
        V_1}^2+\frac{{\widetilde V_1}^2}{2}}& \mbox{ and } & u_{32} = u_2(y_3)-u_2(y_2).
  \end{array}
\end{equation}
 Using similar notations, the extensional stress on both banks of the capacitive transducer
 of port 2  is given by a simpler expression since the bias voltage $V_2$ applied to that port is constant:
\begin{equation}
T_2(y_4)=T_2(y_5)=\frac{\varepsilon_0}{2} \frac{{\overline V_2}^2  }
{\left(g_2+u_{54}\right)^2}
\end{equation} 
with \begin{equation} u_{54} = u_2(y_5)-u_2(y_4).
\end{equation}
At any point along the thickness of the layered structure, the normal
displacement can be split into  static and  dynamic parts:
\begin{equation} 
u_2(y,t)=\bar u_2(y)+\tilde u_2(y,t).
\end{equation}
Furthermore, let us  restrict our attention to the case $\widetilde V_1
\ll \overline V_1$ at the input port. Then we can drop the second harmonic
term ($\cos 2\omega t$) in (\ref{eq:total-drive}). As long as the biasing
voltage stays smaller than the pull--in threshold \cite{Nathanson67} in both gaps,
 causality and visco-elastic damping ruling the propagation
of elastic waves in solids permit us to consider that the dynamic displacement
is small but not negligible with respect to the finite dimensions of the biased structure,
including the small gaps. Then
the dynamic component of the mechanical strain and stress of interest, $S_2$ and $T_2$, respectively, remain
infinitesimal  in the entire structure, so that we can linearize the solution
of the complete  problem
around its static solution:
\begin{equation}\begin{array}{l}
\tilde S_2(y,t) = \lim_{\tilde V_1/\bar V_1\rightarrow 0} \left(S_2(y,t)-\bar
  S_2(y)\right), \vspace{0.6em}\\
\tilde T_2(y,t) = \lim_{\tilde V_1/\bar V_1\rightarrow 0} \left(T_2(y,t)-\bar
  T_2(y)\right).
\end{array}
\end{equation}
Applying this procedure  we obtain the following
first-order expansions of the dynamic extensional stress at all interfaces of
the structure:
\begin{equation}
\label{eq:stresses}
\begin{array}{l}
\widetilde T_2(y_1,t)=0, \vspace{0.6em}\\\displaystyle 
\widetilde T_2(y_2,t)=\widetilde T_2(y_3,t)=\varepsilon_0\left[\frac{ \overline V_1 \widetilde  V_1 \cos \omega t   }
{\left(g_1+\bar u_{32}\right)^2} -  \frac{{
    V_0}^2\, \tilde u_{32}(t)}{\left(g_1+\bar u_{32}\right)^3} \right],
\vspace{0.6em}\\ \displaystyle
\widetilde T_2(y_4,t)=\widetilde T_2(y_5,t)=-\frac{\varepsilon_0  {
    \overline V_2}^2 \,\tilde u_{54}(t)  }
{\left(g_2+\bar u_{54}\right)^3}, \vspace{0.6em}\\\displaystyle 
\widetilde T_2(y_6,t)=0. 
\end{array}
\end{equation}

\section{Solution by Green's function  treatment}
The system of Eqs.~(\ref{eq:stresses}) establishes a useful  relationship between stresses,
normal displacement and electric potential at all interfaces of the
structure. Its linearity permits the use of the complex notation. Denoting the
complex amplitudes of $\widetilde T_2(y_i,t)$ and $\tilde u_2(y_i,t)$ by
$\tau_i$ and $\upsilon_i$, respectively, we easily turn (\ref{eq:stresses})  into a quite systematic form: 
\begin{equation}\label{eq:surface-driving}
\left[ \begin{array}{c}
\tau_1 \\
\tau_2 \\
\tau_3 \\
\tau_4 \\
\tau_5 \\
\tau_6 
\end{array}
\right] = \left[\begin{array}{c} 
0\\
K \widetilde V_1\\
K \widetilde V_1\\
0\\0\\0
  \end{array}
\right]
+  \left[\begin{array}{cccccc} 
0 & 0 & 0 & 0 & 0 & 0 \\
0 & L & -L & 0 & 0 & 0 \\
0 & L & -L & 0 & 0 & 0 \\
0 & 0 & 0 & M & -M & 0 \\
0 & 0 & 0 & M & -M & 0 \\
0 & 0 & 0 & 0 & 0 & 0
\end{array} \right]
\left[ \begin{array}{c}
\upsilon_1 \\
\upsilon_2 \\
\upsilon_3 \\
\upsilon_4 \\
\upsilon_5 \\
\upsilon_6 
\end{array}
\right]
\end{equation}
where:
\begin{equation}
\begin{array}{ccc} \displaystyle
K=\frac{\varepsilon_0 \overline V_1  }
{\left(g_1+\bar u_{32}\right)^2},
&
\displaystyle L=\frac{\varepsilon_0 {V_0}^2 }
{\left(g_1+\bar u_{32}\right)^3},
&
\displaystyle M=\frac{\varepsilon_0 {\overline V_2}^2  }
{\left(g_2+\bar u_{54}\right)^2}.
\end{array}
\end{equation}
The well-known theory of plane  waves propagation along a given axis of a
purely elastic solid  permits to state the
existence of definite transfer matrices for the acoustic propagation in all elastic layers of the structure\cite{Fahmy73}. In
here-considered structure, we only consider cubic or isotropic solids. Then,
the transfer matrices of interest are of dimensions $[2\times2]$,
 as well as the acoustic impedance or admittance matrices and Green's
functions related to them\cite{Smith01}. Thus, the 
admittance form of Green's function matrix\cite{Alex01} of the considered laminar plate forming the
layer $(n)$ is well defined, according to the procedure summarized in the Appendix:
\begin{equation}
\left[\begin{array}{c}
\upsilon_{(2n-1)} \vspace{0.6em}\\ \upsilon_{(2n)}
\end{array}\right]=\left[ \begin{array}{cc}
G_{11}^{(n)} & G_{12}^{(n)}\vspace{0.6em}\\ G_{21}^{(n)} & G_{22}^{(n)} \end{array}\right]
\left[\begin{array}{c}
\tau_{(2n-1)} \vspace{0.6em}\\ \tau_{(2n)}\end{array}\right]
\end{equation}
 Accordingly, a simple assembly of the matrices obtained for the separate
 layers  provides the detailed form of the Green's function matrix characterizing the
 entire  structure:
\begin{equation}\label{eq:surface-impedance}
\left[ \begin{array}{c}
\upsilon_1  \vspace{0.5em}\\
\upsilon_2  \vspace{0.5em}\\
\upsilon_3  \vspace{0.5em}\\
\upsilon_4  \vspace{0.5em}\\
\upsilon_5  \vspace{0.5em}\\
\upsilon_6 
\end{array}
\right] =  \left[\begin{array}{cccccc} 
G^{(1)}_{11} & G^{(1)}_{12} & 0 & 0 & 0 & 0 \vspace{0.4em}\\
G^{(1)}_{21} & G^{(1)}_{22} & 0  & 0 & 0 & 0 \vspace{0.4em}\\
0 & 0 & G^{(2)}_{11} & G^{(2)}_{12} & 0 & 0 \vspace{0.4em}\\
0 & 0 & G^{(2)}_{21} & G^{(2)}_{22} & 0 & 0 \vspace{0.4em}\\
0 & 0 & 0 & 0 & G^{(3)}_{11} & G^{(3)}_{12}\vspace{0.4em} \\
0 & 0 & 0 & 0 & G^{(3)}_{21} & G^{(3)}_{22}
\end{array} \right]
\left[\begin{array}{c}
\tau_1 \vspace{0.5em}\\
\tau_2 \vspace{0.5em}\\
\tau_3 \vspace{0.5em}\\
\tau_4 \vspace{0.5em}\\
\tau_5 \vspace{0.5em}\\
\tau_6
  \end{array}
\right].
\end{equation}
Substituting (\ref{eq:surface-impedance}) into (\ref{eq:surface-driving})
yields a closed--form solution of the problem through basic linear algebra:
\begin{equation}\label{eq:solution}
\vec {\bm \upsilon} = \left[{\bf I}_{(6\times6)}-{\bf G}_{(6\times6)} {\bf P}_{(6\times6)}\right]^{-1}
 {\bf G}_{(6\times6)} \left[\begin{array}{c} 
0\\
K \widetilde V_1\\
K \widetilde V_1\\
0\\0\\0
  \end{array}
\right]
\end{equation}
where ${\bf G}_{(6\times6)}$ is the $6\times6$ matrix appearing in
(\ref{eq:surface-impedance}), ${\bf I}$ denotes the identity matrix, and $
{\bf P} $ is the $6\times6$ matrix appearing in
the right hand member of
 (\ref{eq:surface-driving}).
\section{Electrical response}
\begin{figure}[ht]
\includegraphics[width=0.9\linewidth]{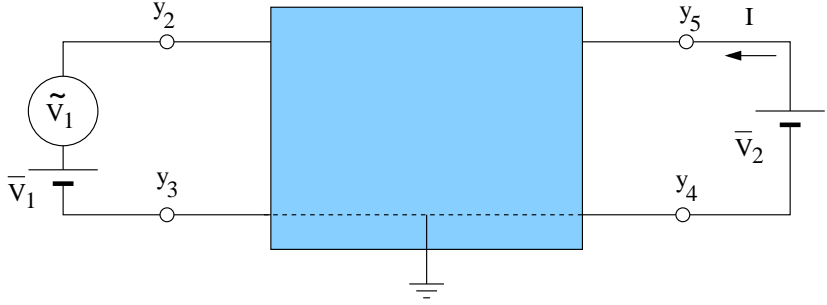}
\caption{Electrical representation of the studied structure.}
\label{fig:quadripole}
\end{figure}
Electromechanical analogies are frequently performed to model the MEMS
resonators via equivalent electrical circuits \cite{Bannon00}. Here-proposed  
approach permits to directly access the electrical response of
thickness--extensional modes in stacked structures.

Let us assume that the silicon layer is electrically grounded.
Then the device is an electrical quadripole as shown on
Fig.~\ref{fig:quadripole}. The  charge borne by the output electrode at $y_5$ is easily
obtained from Gauss' theorem:
\begin{equation}
q=\varepsilon_0\int_S n_2E_2\,dS=-\varepsilon_0\int_S E_2\,dS. 
\end{equation}
The corresponding current delivered by the constant voltage source $\overline
V_2$ to that electrode  is:
\begin{equation}
I=\dot q=\varepsilon_0\overline V_2\int_S\frac{\partial }{\partial
  t}\left(\frac{1}{g_2+\bar u_{54}+\tilde u_{54}}\right)\,dS.
\end{equation}
Performing a Taylor's expansion of the locally--deformed gap with respect
to the small incremental dynamic variation, one obtains
 the corresponding current, with the notations used throughout this paper
\begin{equation} 
I\approx j\omega\varepsilon_0\overline V_2
  \int_S  \frac{\upsilon_4 - \upsilon_5 }{\left(g_2+\bar u_{54}\right)^2}\,dS.
\end{equation}
Then, the current is easily derived from the displacement solution established
in (\ref{eq:solution}).

\section{Application}

\begin{figure}[h]
\includegraphics[width=\linewidth]{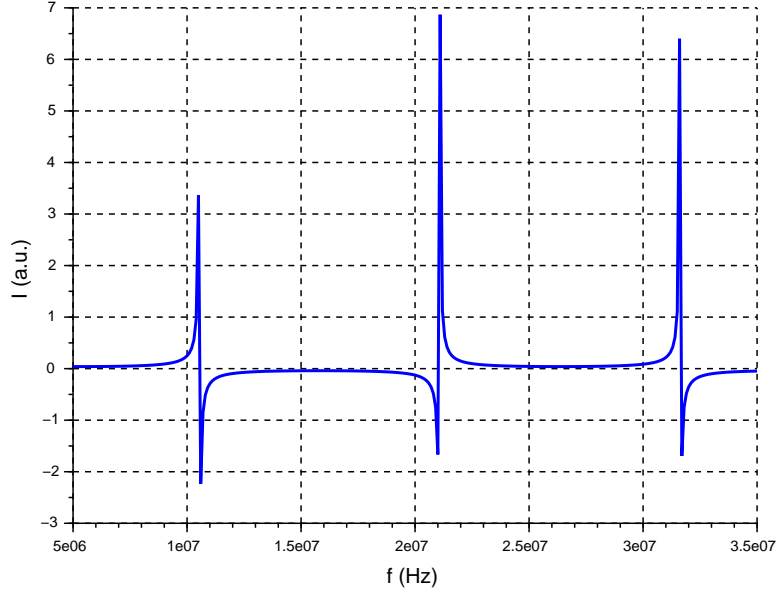}
\caption{Computed imaginary part of current $Im(I)(\tilde V_1)$ in the output port in terms of frequency.}
\label{fig:I-45}
\end{figure}

\begin{figure}[h]
\includegraphics[width=\linewidth]{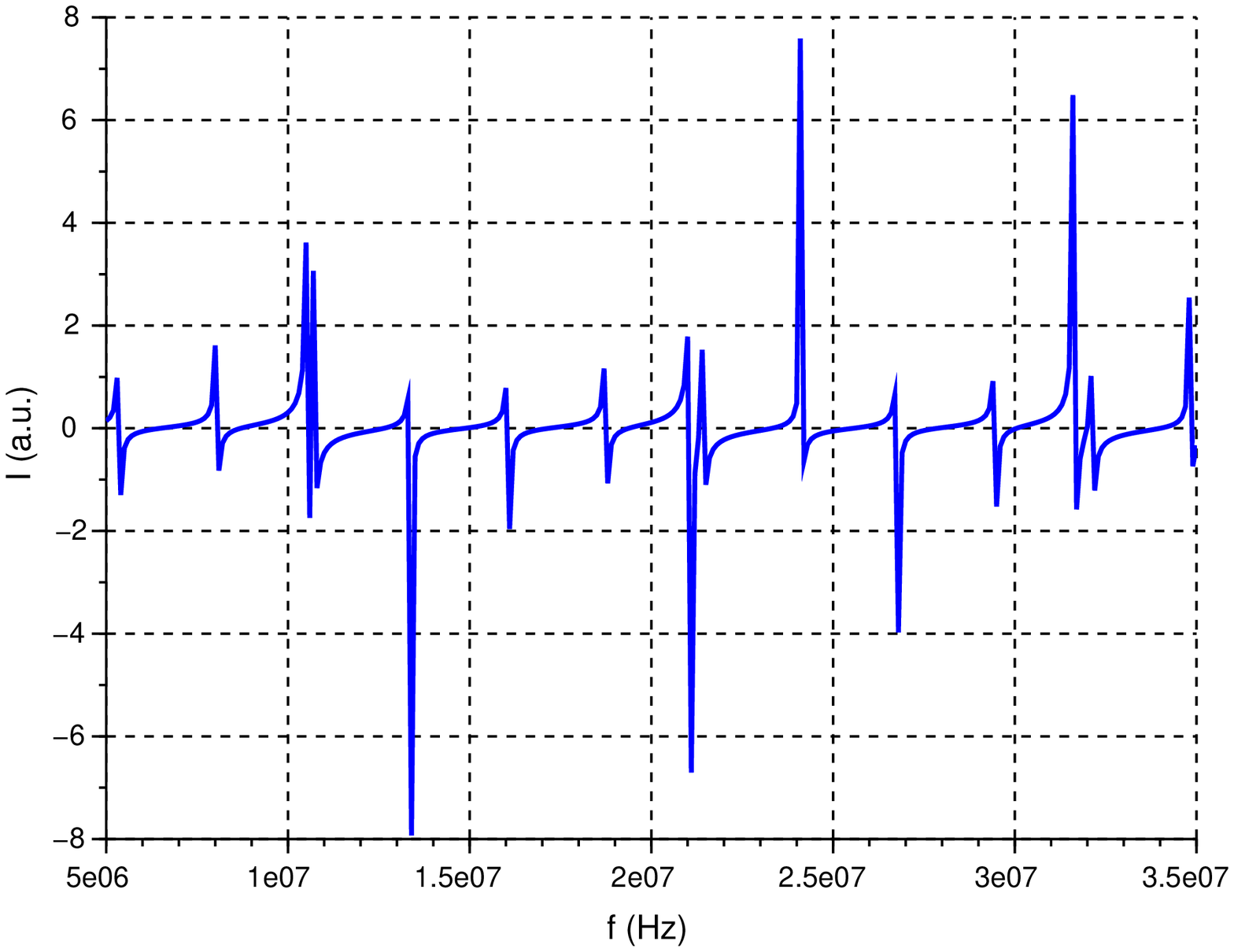}
\caption{Computed imaginary part of the current $Im(I)(\tilde V_1)$ in the input  port in terms of frequency.}
\label{fig:I-23}
\end{figure}

 Fig.~\ref{fig:I-45} shows a plot of the current in the output port in terms of
 frequency, for the following parameters:  $h^{(1)}=h^{(3)}=1\,mm$,
 $h^{(2)}=400\;\mu m$, $\overline V_1=\overline V_2=100\,V$, $\tilde
 V_1=1\,V$.
One observes little influence of the resonances of the bottom layer (glass) onto
the current output. For comparison, we provide on Fig.~\ref{fig:I-23} a plot of
the current delivered by the source at the input port. In that case, the
contribution of vibration of the bottom glass wafer is clearly visible. 
  
Since electrostatic driving is actually performed by an external mechanical
force, either even and odd modes can be driven through the two-ports setup
sown on Fig.~\ref{fig:device}. This also explains that the behavior of the phase observed
on Fig.~\ref{fig:I-45} is different from the behavior that would be observed
upon operating thickness modes in a stack of solid layers with help of some
piezoelectric layer, for instance. This ability to drive both symmetric and
antismmetric modes is easily confirmed by analyzing the
frequency dependence of the vector 
of  mechanical displacements at all interfaces, provided by
Eq.~(\ref{eq:solution}). 

\section*{Acknowldegements}

This work has been partially supported through the MEMS-RF project of the Laboratory of
Excellence FIRST-TF.

\bibliography{biblio}

\begin{thebibliography}{1}

\bibitem{Mason58}
W.~P. Mason.
\newblock {\em Physical {A}coustics and the properties of solid}.
\newblock D. Van Nostrand, Princeton, NJ, 1958.

\bibitem{Nathanson67}
R.~A .~Wickstrom H.~Nathanson, W. E.~Newell and J.~R.~Davis Jr.
\newblock The resonant gate transistor.
\newblock {\em IEEE Trans. Electron Devices}, ED-14:157--176, 1996.

\bibitem{Fahmy73}
A.~H. Fahmy and E.~Adler.
\newblock Propagation of surface acoustic waves in multilayer: a matrix
  description.
\newblock {\em Appl. Phys. Lett.}, (22):495--497, 1973.

\bibitem{Smith01}
P.~M. Smith.
\newblock Dyadic {G}reen's function for multi-layer {SAW} substrates.
\newblock {\em IEEE Trans. on Ultrasonics, Ferroelectrics and Frequency
  Control}, 48(1):171--179, 2001.

\bibitem{Alex01}
A.~Khelif A.~Reinhardt, V.~Laude and S.~Ballandras.
\newblock Dyadic {G}reen's {F}unction of a {L}aminar {P}late.
\newblock {\em IEEE Trans. on Ultrasonics, Ferroelectrics and Frequency
  Control}, 51(9):1157--1163, 2004.

\bibitem{Bannon00}
J.~R.~Clark F.~D.~Bannon, III and T.-C. Nguyen.
\newblock High-{Q} {HF} {M}icroelectromechanical {F}ilters.
\newblock {\em IEEE Journal of Solid-state circuits}, 35(4):512--526, 2000.

\end{thebibliography}
\bibliographystyle{unsrt}

\appendix
\section{Detailed form of required matrices}

A combination of longitudinal plane waves propagating  along both positive and negative directions
of $\vec x_2$ axis in one of the solid layer of the
structure is represented by:
\begin{equation}
\tilde u_2(y)=\left(A^+e^{-j\omega s y} + A^-e^{j\omega s y}\right)\,
e^{j\omega t}
\end{equation}
where $s$ denotes the slowness of the elastic waves in the direction of interest.Then the longitudinal stress created by this elastic wave is given by:
\begin{equation}
\widetilde T_2(y)=-j\omega Z \left(A^+e^{-j\omega s y} - A^-e^{j\omega s y}\right)\,
e^{j\omega t},
\end{equation}
where  $Z=\sqrt{\rho c_{22}}$ is the acoustic impedance of the layer, $c_{22}$
being the elastic constant ruling the propagation along the considered axis,
and $\rho$ is the mass density of the layer.

Let $(n)$ denote this layer. The identification of $\tilde u_2$ and
$\widetilde T_2$
at the coordinate $y_{2n-1}$ in the notations of this paper permits to obtain
$A^+$ and $A^-$. Then a second identification of $\tilde u_2$ and $\widetilde T_2$ at the other
end of the layer, \emph{i.e.}  $y_{2n}$ straightforwardly gives the expression of the transfer
matrix between both surfaces of the said layer. With present notations, the
forward transfer matrix, defined by
\begin{equation}
\left[\begin{array}{c} \upsilon_{(2n)} \vspace{0.4em}\\ \tau_{(2n)}
\end{array}\right] = {\bm M^{(n)}}  \left[\begin{array}{c} \upsilon_{(2n-1)} \vspace{0.4em}\\ \tau_{(2n-1)}
\end{array}\right] 
\end{equation}
has the following expressions:
\begin{equation}  {\bm M^{(n)}}= \left[ \begin{array}{cc} \displaystyle \cos \omega s^{(n)} h_{n} &\displaystyle
  \frac{\sin \omega s^{(n)} h_{n}}{Z^{(n)}\omega} \vspace{0.6em}\\\displaystyle
  -Z^{(n)}\omega \sin  \omega s^{(n)} h_{n} & \displaystyle \cos \omega s^{(n)} h_{n}
\end{array}\right],\vspace{0.4em}
\end{equation}
where the slowness $s^{(n)}$ of the wave in the solid layer $(n)$ is equal to
$\sqrt{\rho/c_{22}^{(n)}}$. The backwards transfer matrix is defined by
\begin{equation}\left[\begin{array}{c} \upsilon_{(2n)} \vspace{0.4em}\\ \tau_{(2n)}
\end{array}\right]
 = {\bm N^{(n)}}  \left[\begin{array}{c} \upsilon_{(2n-1)} \vspace{0.4em}\\ \tau_{(2n-1)}
\end{array}\right] 
\end{equation}
and ${\bm N^{(n)}} =  \left(\bm M^{(n)}\right)^{-1}$ is actually equal to the transpose
of $\bm M^{(n)}$ since  $\mbox{det} ({\bm M})=1$:
\begin{equation}\label{eq:transfer}  {\bm N^{(n)}}= \left[ \begin{array}{cc} \displaystyle \cos \omega s^{(n)} h_{n} &\displaystyle
  -\frac{\sin \omega s^{(n)} h_{n}}{Z^{(n)}\omega} \vspace{0.6em}\\\displaystyle
  Z^{(n)}\omega \sin  \omega s^{(n)} h_{n} & \displaystyle \cos \omega s^{(n)} h_{n}
\end{array}\right].
\end{equation}
The admittance form expression of Green's function matrix of the considered
laminar layer is defined by \cite{Alex01}:
\begin{equation} \left[\begin{array}{c} \upsilon_{(2n-1)} \vspace{0.4em}\\ \upsilon_{(2n)}
\end{array}\right]  
 = {\bm G^{(n)}} \left[\begin{array}{c} \tau_{(2n-1)} \vspace{0.4em}\\ \tau_{(2n)}
\end{array}\right].
\end{equation}
Basic matrix  operations permit to obtain the elements of $\bm G^{(n)}$ from (\ref{eq:transfer}):
\begin{equation}  {\bm G^{(n)}}= \left[ \begin{array}{lc} \displaystyle 
      {\frac{N_{11}^{(n)}}{N_{21}^{(n)}} \hspace{0.6em}} & \displaystyle  N^{(n)}_{12}-\frac{N^{(n)}_{11}N^{(n)}_{22}}{N^{(n)}_{21}}
      \vspace{0.4em}\\\displaystyle 
      \frac{1}{N^{(n)}_{21}} & \displaystyle  -\frac{N^{(n)}_{22}}{N^{(n)}_{21}}
\end{array}\right].
\end{equation}

\vfill

\end{document}